\documentclass[twocolumn,aps,pra,superscriptaddress,amsmath,amssym,floatfix,longbibliography]{revtex4-1}

\usepackage[usenames,dvipsnames]{xcolor}

\usepackage[colorlinks=true,linkcolor=MidnightBlue,citecolor=blue,filecolor=green,urlcolor=PineGreen]{hyperref}
\usepackage{graphicx,epsfig}
\usepackage{color}
\usepackage[usenames,dvipsnames]{xcolor}
\usepackage{amsmath,bbm,amssymb, amsthm,natbib}
\usepackage{dsfont} 
\usepackage{stmaryrd}

\usepackage{numprint}

\begin{document}
	\title{Bosons falling into a black hole: A superfluid analogue} \author{Sreenath K. Manikandan}
	\email{skizhakk@ur.rochester.edu}
	\affiliation{Department of Physics and Astronomy, University of Rochester, Rochester, NY 14627, USA}
	\affiliation{Center for Coherence and Quantum Optics, University of Rochester, Rochester, NY 14627, USA}   
	\author{Andrew N. Jordan}
	\email{jordan@pas.rochester.edu}
	\affiliation{Department of Physics and Astronomy, University of Rochester, Rochester, NY 14627, USA}
	\affiliation{Center for Coherence and Quantum Optics, University of Rochester, Rochester, NY 14627, USA}    
	\affiliation{Center for Quantum Studies, Chapman University, Orange, CA, USA, 92866}
	\date{\today}
	
	\begin{abstract}
We propose an analogy between the quantum physics of a black hole in its late stages of the evaporation process and a superfluid Bose Einstein Condensate (BEC), based on the Horowitz and Maldacena quantum final state projection model [JHEP 2004(02), 008]. The superfluid region is considered to be analogous to the interior of a black hole, and the normal fluid/superfluid interface is compared to the event horizon of a black hole. We theoretically investigate the possibility of recovering the wavefunction of particles incident on a superfluid BEC from the normal fluid, facilitated by the mode conversion processes occurring at the normal fluid/superfluid BEC interface. We also study how the correlations of an infalling mode with an external memory system can be preserved in the process, similar to Hayden and Preskill's ``information mirror'' model for a black hole [JHEP 2007(09), 120]. Based on these analogies, we conjecture that the quantum state of bosons entering a black hole in its final state is the superfluid quantum ground state of interacting bosons. Our analogy suggests that the wavefunction of bosons falling into a black hole can be recovered from the outgoing Hawking modes. In the particular case when a hole-like quasiparticle (a density dip) is incident on the superfluid BEC causing the superfluid to shrink in size, our model indicates that the evaporation is unitary.
	\end{abstract}
	
	\maketitle
	\section{Introduction}
Certain many body quantum wavefunctions can be considered to be rigid, in the sense that they can act like fixed points in a Hilbert space, where the quantum dynamics around these states would exhibit interesting physical properties without changing the wavefunction of the rigid region. Black holes are thought to be one such physical system where this is true. In fact such an assumption resolves the black hole information problem since certain final states permit teleporting the information contained in the infalling matter to outgoing Hawking radiation~\cite{horowitz2004black,gottesman2004comment,lloyd2014unitarity}. The black hole described by these models is in its late stages of the evaporation process, past its ``half-way point" where more than half of its initial entropy has been radiated away~\cite{hayden2007black}, such that an asymptotic quantum final state can be envisaged. Recently, the authors have proposed that albeit being a completely different physical system, superconductors have many of the same features of a black hole, and the final state projection model applies to analogous Hawking radiation (Andreev reflection) from a superconductor, thereby mapping several existing theoretical resolutions of black hole information puzzle to experiments using superconductors~\cite{manikandan2017andreev}.
        
Here we extend this analogy to the case of bosonic superfluids.  We consider a normal fluid/superfluid interface similar to the experimental setup discussed by Zapata and Sols~\cite{zapata2009andreev} to describe a bosonic version of Andreev reflections~\cite{andreev1964thermal}. In the bosonic analogue, a particle-like mode (density bump) incident on the superfluid from the normal side, triggers the retro-reflection of a hole-like quasiparticle (density dip), which propagates in the normal region with a group velocity that is different from ordinary reflections. In experiments, the distinction between normal fluid and superfluid can be made with respect to the local speed of sound; The normal region is where the fluid flows faster than the local speed of sound. Naively, a bosonic quasiparticle incident on the superfluid from the normal fluid gets ``trapped" in the superfluid because in order to return to the normal region, it has to achieve velocities exceeding the local speed of sound.  Such a situation can be constructed by considering a decaying condensate, with the outgoing coherent beam treated as the normal region where the condensate interactions can be ignored~\cite{leboeuf2001bose,zapata2009andreev}. Here, the boundary between the decaying condensate and the outgoing coherent beam plays the role of an event horizon, similar to other sonic black hole  analogies~\cite{visser1998acoustic,lahav2010realization,jannes2009emergent,garay2000sonic,giovanazzi2005hawking} and black hole lasers~\cite{steinhauer2014observation}.

Our approach can be contrasted with existing proposals that compare superfluid BECs to black holes, which aim at resolving various aspects of the particle production problem near a black hole, pioneered by Prof. Stephen Hawking~\cite{hawking1975particle,gibbons1977cosmological}. The first paper which mention an analogy between Andreev reflections and Hawking radiation was authored by Prof. Jacobson~\cite{jacobson1996origin}, where his attempt was to account for the outgoing black hole modes as a result of mode conversion from ingoing modes, and not produced by a Trans-Plankian reservoir at the event horizon. In his 1996 publication, Prof. Jacobson had concluded that a mechanism to recover the quantum information lost in black holes remains elusive, since the partner modes are still lost in the abyss of a black hole when Hawking modes are produced. We combined his observations with the final state projection model proposed by Horowitz and Maldacena, and discuss how the information puzzle can also be studied in this analogy, therefore showing that it offer resolutions to both the Trans-Plankian problem and the information puzzle at once.
\begin{figure}		\includegraphics[scale=0.23]{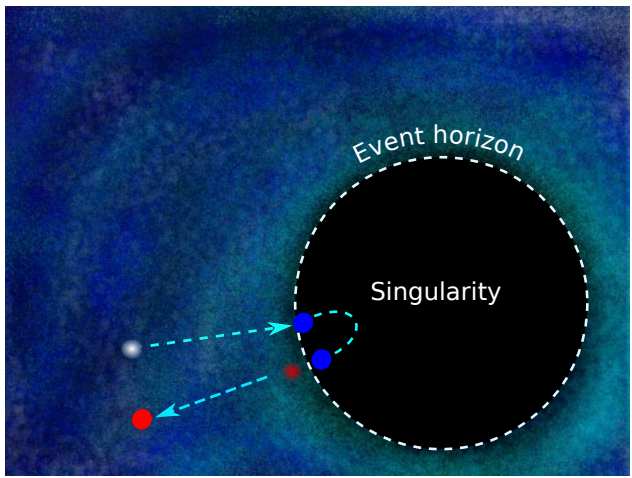}	\caption{Quantum information dynamics near the event horizon of a black hole in Horowitz-Maldacena final state projection model~\cite{horowitz2004black,lloyd2014unitarity}. The incoming quasiparticle mode falls into the black hole by taking the negative energy particle from the Hawking pair, and forming a two mode paired entangled state inside the event horizon, while the outgoing Hawking mode escapes to infinity. In the process, the quantum information originally encoded in the incoming mode is dynamically transferred to the outgoing mode.\label{fig1}}
		\end{figure}

Recently there has been much interest in the superfluid/black hole analogy, for example, the proposal by Zapata et al~\cite{zapata2011resonant}., also studies bosonic Andreev reflection as analogous to Hawking radiation from black holes, although their work mainly focused on characterizing Hawking radiation from atomic condensates with a subsonic/supersonic interface, and not the quantum information paradox aspect of the problem. Similar sonic analogies to black holes using BECs have been proposed before~\cite{lahav2010realization,jannes2009emergent,garay2000sonic}, and a second order superfluid phase transition near the horizon similar to Helium$-4$ has been discussed by Hennigar et al~\cite{hennigar2017superfluid}. Thermodynamic features of an interacting Bose gas is also a topic of major interest in this context, for example,  area law for the scaling of entropy has already been proposed for an interacting system of Helium$-4$~\cite{herdman2017entanglement}. Here, we further investigate the superfluid/ black hole analogy from an information theory point of view for bosons, and show that the unitarity of black hole evaporation in final state projection model is also manifest in mode conversion processes occurring at normal fluid/superfluid BEC interfaces. 

Compared to existing approaches describing mode conversion processes at bosonic superfluid/ normal fluid interface~\cite{zapata2011resonant,zapata2009andreev}, our approach differs in that we assume a microscopic quantum description of the superfluid region as the ground state of interacting bosons~\cite{guo2017berry,fetter2012quantum}. We then look at quasiparticle excitations incident on the superfluid from the normal region, produced from fluctuations of the condensate/relative flow of the condensate, and quasiparticles introduced externally. These quasiparticles are assumed to be sufficiently low in energy such that they do not perturb the condensate away from its ground state. Considering an interacting Bose gas instead of a non-interacting Bose gas permits us to investigate resolutions to information paradox in black holes using final state projection models~\cite{horowitz2004black,lloyd2014unitarity}, since different bosonic modes in the ground state of an interacting Bose gas exist as entangled pairs~\cite{fetter2012quantum,guo2017berry,vedral2003entanglement}. Further, the condensate wavefunction also has certain invariance (under the transformation $\textbf{k}\rightarrow -\textbf{k}$) when the superfluid interactions impose a strict \textit{s-wave} pairing symmetry, which is crucial for the present study.

The closest publication to the current work we are aware of is the fermionic superfluid/black hole analogy presented in~\cite{manikandan2017andreev}. Similar to the fermionic case, we find that several existing proposals which describe black hole evaporation as a unitary process has analogies in the bosonic case as well, notably, the models proposed by Horowitz and Maldacena~\cite{horowitz2004black}, and Preskill and Hayden~\cite{hayden2007black}. Based on this analogy, we propose a final quantum state for bosons falling into a black hole, which is the ground state wavefunction of a superfluid BEC where particles exist as entangled pairs. This can be viewed in par with the previously proposed superconducting BCS wavefunction of fermions forming a black hole~\cite{manikandan2017andreev} where they exist in the form of Cooper  pairs~\cite{bardeen1957theory}, and therefore gives a complete symmetrical picture, prescribing wavefunctions to fermions and bosons inside a black hole. The advantage of our approach is that black hole evaporation can be seen to be a manifestly unitary process when the interactions are maximally entangling, and it resolves the black hole information problem for bosons, similar to resolutions of the black hole information paradox for fermions proposed previously~\cite{manikandan2017andreev}. The notion of a black hole in our present study is also consistent with the ``information mirror" model proposed by Hayden and Preskill~\cite{hayden2007black}, and the black hole/superconductor analogy previously proposed by the authors, where the black hole `Andreev' reflects the quantum information encoded in the collapsing matter, while accepting particles~\cite{manikandan2017andreev}.

A detailed analysis of various presumptions leading to the condensate/black hole analogy has already been presented elsewhere~\cite{manikandan2017andreev}. To be brief, our generic approach to resolving the information paradox using condensates require that we assume the validity of a local quantum field theoretic description near the event horizon of a black hole, such that one can factorize the Hilbert space into subspaces, and define correlations such as entanglement. As has been discussed previously, such an assumption, together with the assumption of linearity and unitarity of the quantum theory require that the wavefunction of the interior of the black hole contain no information about future/ past Cauchy surfaces, and is special~\cite{susskind1993stretched,chakraborty2017black,manikandan2017andreev}. Our effort is to show, using information considerations, that condensate wavefunctions have many features to qualify as this special quantum state to describe the interior of a black hole, for fermions~\cite{manikandan2017andreev} and bosons (present study) respectively.

This article is organized as follows: In Sec.~\ref{sec1} we show that the quantum information encoded in continuous variable modes falling into a superfluid BEC can be recovered from the outgoing modes, and the process can be understood as a deterministic continuous variable quantum teleportation~\cite{vaidman1994teleportation} of information, mediated by local tunneling/pairing interactions. The implications to the final quantum state of an evaporating black hole is discussed in Sec.~\ref{sec3}, by making analogy with the Horowitz-Maldacena final quantum state proposal for black holes~\cite{horowitz2004black}. The analogy with the black hole ``information mirror" model proposed by Hayden and Preskill~\cite{hayden2007black} is also discussed in Sec.~\ref{preskill}, where we show that the condensate preserves quantum correlations of an infalling mode with an external memory system by transferring them to the outgoing mode. We discuss our main results and conclusions in Sec.~\ref{conclusion}.
	\section{Mode conversion at the interface between a normal fluid and a superfluid BEC\label{sec1}}
In this section we describe the mode conversion processes at the interface between a normal fluid and a superfluid BEC of interacting bosons, when a particle like mode (density bump) is incident on the superfluid from the normal side. In practice, a few processes result~\cite{zapata2009andreev}: (1) ordinary reflection, where the particle like mode reflects from the interface back to the normal region, (2) transmission of the quasi-particle across the condensate, and (3) Andreev reflection of the quasiparticle, where the retro-reflected quasiparticle is hole-like, observed as a density dip. While all these processes are of interest for various applications of quasiparticle transport across normal fluid/superfluid interfaces~\cite{zapata2009andreev}, we restrict to situations where the interface is ideal (low reflectivity) and vanishingly small excitation energies such that processes (1) and (2) can be respectively ignored. The time reversals of these processes are also physical, for example, the time reversal of (3) would be the case when a hole-like quasiparticle (density dip) incident on the superfluid from the normal side trigger a particle-like mode (density bump) to be emitted from the superfluid to the normal side, causing the superfluid to shrink in size.

\begin{figure*}		\includegraphics[scale=0.25]{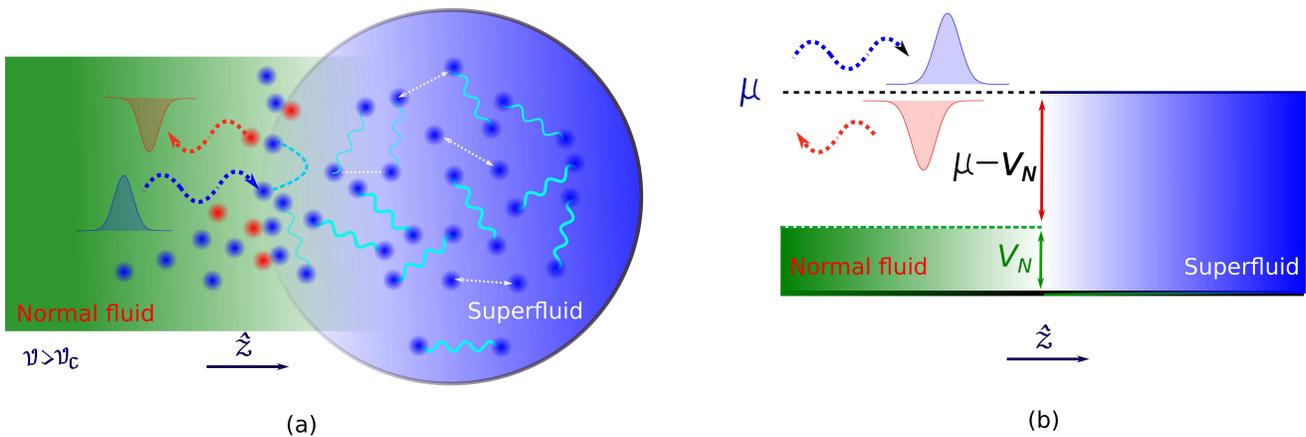}	\caption{ (a) Boundary between a normal fluid (the fluid velocity $v>v_{c}$, where $v_{c}$ is the local speed of sound in the superfluid region) and the superfluid. In the low energy ($\varepsilon\rightarrow 0$) limit, a coherently populated incoming quasiparticle mode (density bump, excitation energy $+\varepsilon$, blue) can enter the superfluid by taking the negative energy (-$\varepsilon$) quasi-particle from the bosonic particle hole pair available at the interface, and forming a two mode squeezed state in the superfluid. The hole like partner mode (absence of the negative energy ($-\varepsilon$) quasi-particle, and hence a density dip, red) now populated coherently, escapes to infinity. It has energy $+\varepsilon$ from particle-hole symmetry. Interactions in the superfluid state are also shown, where scattering events dynamically create (and break) entanglement between different modes in the condensate. (b) Andreev scattering diagram for a leaking condensate ($\mu-V_{N}>0$) for vanishingly small excitation energy $\varepsilon$~\cite{zapata2009andreev}. \label{fig2}}
		\end{figure*}

We begin by reviewing a simple model for the BEC of interacting bosons by closely following the discussions in~\cite{guo2017berry,fetter2012quantum}. The Hamiltonian for an interacting Bose gas whose ground state is the superfluid BEC is given by (in the units $\hbar=m=1$),
\begin{eqnarray}
H=\sum_{\textbf{p}}\varepsilon_{\textbf{p}}d_{\textbf{p}}^{\dagger}d_{\textbf{p}}+\frac{g}{2V}\sum_{\textbf{p},\textbf{p}'}d_{\textbf{p}}^{\dagger}d_{-\textbf{p}}^{\dagger}d_{\textbf{p}'}d_{-\textbf{p}'},
\end{eqnarray}	
where $\varepsilon_{\textbf{p}}=\frac{|\textbf{p}|^{2}}{2}$, $g$ is the coupling constant and $V$ is the volume. In the limit when $\frac{N-N_{0}}{N}\ll 1$, where $N$ is total number of particles and $N_{0}$ is the average number of particles in the ground state, we can write down the following effective Hamiltonian for the BEC~\cite{guo2017berry,fetter2012quantum}:
\begin{eqnarray}
H_{\text{BEC}} &=& \frac{1}{2}vgn^2+\frac{1}{2}\sum_{\textbf{p}\neq 0}[(\varepsilon_{\textbf{p}}+ng) (d_{\textbf{p}}^{\dagger}d_{\textbf{p}}+d_{-\textbf{p}}^{\dagger}d_{-\textbf{p}})\nonumber\\&+&ng(e^{-2i\tau}d_{\textbf{p}}^{\dagger}d_{-\textbf{p}}^{\dagger}+\text{H.c.})],
\end{eqnarray}
where $n=\frac{N}{V}$ and $\tau$ is the phase of the condensate.  This Hamiltonian can be diagonalized using Bogoliubov transformations and the superfluid ground state can be computed as the Bogoliubov quasiparticle vacuum,
\begin{equation}
|\psi_{\text{BEC}}\rangle = e^{\frac{1}{2}\sum_{\textbf{p}}\theta_{\textbf{p}}(d_{\textbf{p}}d_{-\textbf{p}}e^{2i\tau}-d_{\textbf{p}}^{\dagger}d_{-\textbf{p}}^{\dagger}e^{-2i\tau})}|0\rangle,\label{BEC}
\end{equation}	
where $\cosh(2\theta_{\textbf{p}})=\frac{\varepsilon_{\textbf{p}}+ng}{\sqrt{(\varepsilon_{\textbf{p}}+ng)^{2}-(ng)^{2}}}$, and $\sinh(2\theta_{\textbf{p}})=\frac{ng}{\sqrt{(\varepsilon_{\textbf{p}}+ng)^{2}-(ng)^{2}}}$~\cite{guo2017berry,fetter2012quantum}.

The crucial observation for the present study is that the ground state wavefunction of a superfluid BEC given in Eq.~\eqref{BEC} is a coherent squeezed state, which exists as pairs of particles in opposite momentum quantum states~\cite{guo2017berry,fetter2012quantum}. For the remainder of the section, we consider one such pair with squeezing parameter $r$ and momentum $\textbf{p}$, and denote it by $|\psi(r,\textbf{p})\rangle$:
\begin{eqnarray}
|\psi(r,\textbf{p})\rangle &=& \exp(re^{i\phi}~d_{\textbf{p}}^{\dagger}d_{-\textbf{p}}^{\dagger}-re^{-i\phi}~d_{\textbf{p}}d_{-\textbf{p}})|0_{\textbf{p}}0_{-\textbf{p}}\rangle\nonumber\\&=&\text{sech}(r)\sum_{n=0}^{\infty}[e^{i\phi}\tanh(r)]^{n}|n_{\textbf{p}}\rangle_{d}|n_{-\textbf{p}}\rangle_{d},
\end{eqnarray}
where $\phi$ can be thought of as the condensate phase. This state is highly entangled. The subscript `$d$' is used to distinguish modes inside the condensate/ falling into the condensate from the outgoing modes, labeled by `$a$'. The average quanta in each mode is $\langle n_{\textbf{p}}\rangle=\langle n_{-\textbf{p}}\rangle=\sinh(r)^2$. The statistics of each mode is also a pseudo-thermal state with an effective temperature T defined by $\exp(-\beta\omega) = \tanh(r)^2$, where $\beta=\frac{1}{k_{B}\text{T}}$~\cite{guo2017berry,vedral2003entanglement}. In the following, we show that existence of particles only in the form of entangled pairs within the condensate requires that the quantum information encoded in the wavefunction of particles falling into the condensate cannot enter the condensate, but has to be retro-reflected out of the condensate via mode conversion processes occurring at the interface.

This can be understood as follows. In bosonic Andreev reflections, a quasiparticle (particle like, density bump), incident on the superfluid boundary from the normal fluid can be retro-reflected as a hole-like quasiparticle (density dip) to the normal region, while both of them contribute positively to the density of the condensate~\cite{zapata2009andreev}. In our model, this is described by a quasiparticle mode (particle like, density bump) from a (bosonic) particle/hole pair falling into the condensate with the incoming quasiparticle (particle like, a density bump) and forming a paired coherent state of BEC inside the condensate while leaving behind hole-like quasiparticle (density dip) which propagate back to the normal region, as depicted in Fig.~\ref{fig2}. The converse of this process is also physical, and leads to particle like modes emitted from the condensate upon incidence of a low density wave from the normal fluid. 
        
A dynamic picture of various scattering events in the superfluid state which create and annihilate pairs of particles is depicted in Fig.~\ref{fig2}~(a), together with the scattering events at the boundary which can add/remove pairs in the condensate. Note the similarity of bosonic Andreev processes at the interface between a superfluid BEC and the normal fluid with their fermionic counter part, which is also identified as the microscopic origin of proximity effect in superconductors~\cite{klapwijk2004proximity}: the superconducting correlations can extend to the normal metal in a superconductor/normal metal junction, up to a length scale equal to the coherence length of the superconducting condensate.  In the bosonic case, the corresponding length scale over which the condensate correlations smear out is known the ``healing length" of the condensate, given by \begin{equation}\lambda = \frac{1}{\sqrt{8\pi n a}} = \frac{\hbar}{\sqrt{2}mv_{c}},\end{equation}
where $a$ is the \textit{s-wave} scattering length of the condensate and $v_{c}$ is the speed of sound in the superfluid~\cite{dalfovo1999theory}.

 To lowest non-vanishing order, one can identify the chemical potential $\mu$ as $\mu=\frac{d\varepsilon_{0}}{dN} = ng$, where $\varepsilon_{0}$ is the ground state energy: $\varepsilon_{0}=\langle\psi_{\text{BEC}}|H_{\text{BEC}}|\psi_{\text{BEC}}\rangle$~\cite{fetter2012quantum}. The chemical potential of the superfluid ground state also agrees with the asymptotic value of the chemical potential describing the condensate in Zapata and Sol's approach~\cite{zapata2009andreev}. At the interface with a normal fluid, we assume that the condensate interaction strength $g$ is gradually reduced, and to negligible magnitudes in the outgoing coherent beam (normal fluid). This facilitates pairing of the incoming mode -- having energy slightly above the asymptotic chemical potential of the condensate -- with an infalling mode -- having energy slightly below the asymptotic chemical potential of the condensate (still higher than the local chemical potential, as required for bosons) --   at the interface.
 
In analogy with the fermionic case, a particle like mode represents propagation of a bosonic quasiparticle excitation above the chemical potential of the condensate (above the Fermi level, in the fermionic case) on the normal side, while a hole like mode represents propagation below the chemical potential (below the Fermi level, in the fermionic case) of the condensate on the normal side. Following the discussion in~\cite{zapata2009andreev}, the normal side is described by a flat potential $V_{N}$ for simplicity [see Fig.~\ref{fig2}~(b)]. As has been pointed out previously~\cite{zapata2009andreev}, a decaying condensate will have the chemical potential slightly above the potential $V_{N}$ on the normal side ($\mu>V_{N}$), and hence can permit propagating hole-like modes on the normal side below the asymptotic condensate chemical potential, unlike the case of a confined condensate ($V_{N}>\mu$) where the hole like modes can only be evanescent on the normal side~\cite{zapata2009andreev}. In the following, we assume that the participating modes have energies within the allowed crossing range, $\varepsilon~\epsilon~(V_{N}-\mu,~\mu-V_{N})$ where we consider $\varepsilon$ to be very small such that the energy of the incident particles do not exceed the smallest excitation energy of the condensate. Within this assumption, we can safely assume that the condensate remains in a stationary state, which is the ground state of the interacting bosons. Our assumption is also in agreement with a statistical description of the condensate, where $\varepsilon$ being small correspond to the energy range $(V_{N}-\mu,~\mu-V_{N})$ near the chemical potential, and hence represent quasiparticles that can enter/ or leave an effective grand-canonical ensemble that describes the superfluid BEC.

The interface between the normal fluid and the superfluid BEC is rather special since it involves fluctuations of the condensate. We restrict to a Hamiltonian description of the interface with the Hamiltonian that describes generation of Bogoliubov quasiparticles at the interface,
\begin{equation}
H_{I} = iJ_{\textbf{k},-\textbf{k}}~d_{\textbf{k}}^\dagger a_{-\textbf{k}}^{\dagger}+\text{H.c.},\label{hint}
\end{equation}
where $d_{\textbf{k}}^{\dagger}$ represent creation of an infalling quasiparticle mode with negative energy $-\varepsilon$ having wavevector $\textbf{k} = -\textbf{k}_{\mu}+\delta \textbf{k}$, where $|\textbf{k}_{\mu}|= \sqrt{2(\mu-V_{N})}$. The mode $ a_{-\textbf{k}}^{\dagger}$ describes a simultaneously created low density mode with wavevector $\textbf{k}_{\mu}-\delta \textbf{k}$  on the normal side such that the particle density is conserved globally. Here $iJ_{\textbf{k},-\textbf{k}}$ are effective matrix elements which describe the quasiparticle tunneling across the interface. The resonant interaction which creates/destroys a propagating mode of lower density (hole-like) on the normal side is appropriate for a leaking condensate, for which the chemical potential $\mu$ is slightly above the potential on the normal side ($\mu>V_{N}$), since they permit hole-like modes on the normal side which are not evanescent~\cite{zapata2009andreev}. The Hamiltonian in Eq.~\eqref{hint} is also similar to the tunneling Hamiltonian considered in the fermionic case~\cite{manikandan2017andreev}, where we notice that the interaction creates resonances across the interface where particle mode tends to be on either side of the interface.  Such bosonic quasiparticle/hole pairs are highly correlated. To see this, we look at the time evolution of the vacuum $|0_{\textbf{k}}\rangle_{d}|0_{-\textbf{k}}\rangle_{a}$ by this interaction. We find that,
\begin{eqnarray}
e^{-iH_{I}t}|0_{\textbf{k}}\rangle_{d}|0_{-\textbf{k}}\rangle_{a} &=& e^{tJ_{\textbf{k},-\textbf{k}} d_{\textbf{k}}^\dagger a_{-\textbf{k}}^{\dagger}-tJ_{\textbf{k}} d_{\textbf{k}} a_{-\textbf{k}}}|0_{\textbf{k}}\rangle_{d}|0_{-\textbf{k}}\rangle_{a} \nonumber\\&=& \hat{S}(\zeta)|0_{\textbf{k}}\rangle_{d}|0_{-\textbf{k}}\rangle_{a},
\end{eqnarray}
where $\hat{S}(\zeta) = e^{\zeta d_{\textbf{k}}^\dagger a_{-\textbf{k}}^{\dagger}-\zeta^{*} d_{\textbf{k}} a_{-\textbf{k}}}$ is the two-mode squeezing operator with squeezing parameter $\zeta = J_{\textbf{k},-\textbf{k}}t$, which we assume to be a real number. Although we consider a simple unitary description of particle/hole production based on quasiparticle tunneling at the interface, our approach effectively describes bosonic quasiparticle production at sonic horizons by relative motion of fluids, previously studied in similar contexts~\cite{unruh1981experimental,visser1998acoustic,lahav2010realization}. Further note that the resulting unitary evolution is also consistent with the Bogoliubov representation of the BEC fluctuations in terms of the infalling and outgoing modes. To see this, we look at the time evolution of the modes $d$ and $a$ in the Heisenberg picture. We find that,
$
d_{\textbf{k}}(\zeta) = d_{\textbf{k}}(0)\cosh{\zeta}+a_{-\textbf{k}}(0)^{\dagger}\sinh{\zeta}$ and $a_{-\textbf{k}}(\zeta) = a_{-\textbf{k}}(0)\cosh{\zeta}+d_{\textbf{k}}(0)^{\dagger}\sinh{\zeta}$,
which are Bogoliubov transformations of the bosonic modes. 

We now consider a simple case when the infalling bosonic mode labeled by wavevector $\textbf{q}= \textbf{k}_{\mu}+\delta \textbf{k}$, and positive excitation energy $\varepsilon$ (relative to $\mu$), in the state $|\psi_{in}\rangle_{\textbf{q}}$, with the mode $\textbf{q}$ coherently populated:
\begin{equation}
|\psi_{in}\rangle_{\textbf{q}}=|\alpha_{\textbf{q}}\rangle=e^{-|\alpha|^{2}/2}\sum_{n}\frac{\alpha^{n}}{\sqrt{n!}}|n_{\textbf{q}}\rangle_{d},
\end{equation}
where we denote $|n_{\textbf{q}}\rangle_{d}=\frac{(d_{\textbf{q}}^{\dagger})^{n}}{\sqrt{n!}}|0_{\textbf{q}}\rangle_{d}$ describing $n$ quanta in an infalling mode with wavevector $\textbf{q}$. The choice $|\alpha|=1$ ensures that the average energy of the infalling mode is clearly within the allowed crossing range for the example considered. We can write the joint state at the interface as a tensor product of the incoming state $|\psi_{in}\rangle_{\textbf{q}}$, and the particle hole pair labeled by the particle wavevector $\textbf{k}$:
\begin{equation}
|\Psi\rangle =  e^{-|\alpha|^{2}/2}\sum_{n}\frac{\alpha^{n}}{\sqrt{n!}}|n_{\textbf{q}}\rangle_{d}\otimes\hat{S}(\zeta)|0_{\textbf{k}}\rangle_{d}|0_{-\textbf{k}}\rangle_{a}.
\end{equation} 
Since the condensate only permits pairs of entangled particles to enter, the reduced state of the outgoing mode is found by applying the projection onto the state $|\psi(r,\textbf{k}_{\mu})\rangle$ for the infalling modes. We consider the limit of vanishingly small $\varepsilon$ ($\delta \textbf{k}\rightarrow 0$), which corresponds to the wave vector of particles which can exist both in the condensate and the normal fluid. We obtain:
\begin{eqnarray}
&&\langle\psi|\Psi\rangle\simeq \text{sech}(r)\sum_{n''=0}^{\infty}[e^{-i\phi}\tanh(r)]^{n''}\langle n''_{\textbf{k}_{\mu}}|_{d}\langle n''_{-\textbf{k}_{\mu}}|_{d}e^{\frac{-|\alpha|^{2}}{2}}\nonumber\\&&\times\sum_{n'}\frac{\alpha^{n'}}{\sqrt{n'!}}|n'_{\textbf{k}_{\mu}}\rangle_{d}\otimes\text{sech}(\zeta)\sum_{n=0}^{\infty}\tanh(\zeta)^{n}|n_{-\textbf{k}_{\mu}}\rangle_{d}|
n_{\textbf{k}_{\mu}}\rangle_{a}\nonumber\\&&
=\text{sech}(r)\text{sech}(\zeta)e^{\frac{-|\alpha|^{2}}{2}}\sum_{n=0}^{\infty}\frac{[e^{-i\phi}\tanh(r)\tanh(\zeta)\alpha]^{n}}{\sqrt{n!}}|n_{\textbf{k}_{\mu}}\rangle_{a}\nonumber\\&&\propto|\alpha e^{-i\phi}\tanh(r)\tanh(\zeta)\rangle_{a}.\label{StTrans}
\end{eqnarray}
When $r,~\zeta\gg 1$, equivalently when the interaction time $t\gg\text{max}[J_{\textbf{k},-\textbf{k}}^{-1},~(ng)^{-1}]$, we have the outgoing mode in the coherent state $|\alpha e^{-i\phi}\rangle$, where $\phi$ can be associated to the phase of the condensate. This addition of a relative phase upon Andreev reflection is known for Andreev reflection of an electron (hole) at normal metal/superconductor interfaces~\cite{andreev1964thermal,pannetier2000andreev,klapwijk2004proximity}, and the final state projection approach also predicts the same phase difference between the incoming and outgoing modes~\cite{manikandan2017andreev}. The relative phase between ingoing and outgoing modes in bosonic Andreev reflections as described above can also be measured, and therefore can be verified experimentally.  

In the general case (for an arbitrary incoming quantum state $|\psi_{in}\rangle$), we note that the quantum state of incoming and outgoing modes in Eq.~\eqref{StTrans} are related by the `scattering' operator $\mathcal{U}$: \begin{equation}|\psi_{f}\rangle\propto\mathcal{U}|\psi_{in}\rangle=[e^{-i\phi}\tanh(r)\tanh(\zeta)]^{\hat{N}}|\psi_{in}\rangle,\end{equation} 
where we have suppressed the mode indices indicating the conversion of modes. For identical effective temperature of the final state of the infalling modes and the shared entangled pair at the interface (i.e. when $r=\zeta$), note that the operator $\mathcal{U}$ relates directly to the effective temperature by the relation, $\exp(-\beta\omega) = \tanh(r)^2$ as,
\begin{equation}\mathcal{U}=[e^{-i\phi}~\exp(-\beta\omega)]^{\hat{N}} = \exp(-\hat{N}\beta\omega)~e^{-i\hat{N}\phi}.\label{EqU}\end{equation}
This suggests that the unitarity in final state projection holds only when the pairs in the superfluid BEC are maximally correlated (the superfluid interactions are maximally entangling), i.e. when $r,~\zeta\gg 1$. Nevertheless, the hyperbolic tangent function $\tanh(x)\rightarrow 1$ rather fast w.r.t $x$; a simple accounting in terms of the tunneling/ superfluid interactions suggests that this holds when the interaction time $t\gg\text{max}[J_{\textbf{k},-\textbf{k}}^{-1},~(ng)^{-1}]$. Thus, at relatively large time scales relevant for the scattering problem, the scattering matrix $\mathcal{U}$ becomes unitary in the final state projection approach.
We emphasize that the apparent departure from unitarity in the final state projection approach when applied to short time scales does not imply breakdown of unitarity of physics at the BEC interface. The seemingly lost amplitude in this effective model would correspond to the probability amplitude to remain unpaired, and leakage of the incoming quasiparticle amplitude through other boundary processes which the final state projection approach does not account for, such as ordinary reflection, and transmission of the quasiparticle across the condensate.

Notice that the limit $r,~\zeta\gg 1$ is a bit tricky since the hyperbolic secant function vanish in the limit. We can better understand the limit $r,~\zeta\rightarrow \infty$ using a Heisenberg picture description, by following some simple arguments in the language of continuous variable quantum teleportation~\cite{vaidman1994teleportation}: In the Heisenberg picture, the particle hole pair of bosons produced at the interface between the normal fluid and the superfluid BEC should have all the symmetries of the vacuum, and hence they should exist in time reversed states of each other:
\begin{equation}
\hat{x}_{\textbf{k}}=\hat{x}_{\textbf{h}}~~\text{and}~~\hat{p}_{\textbf{k}}=-\hat{p}_{\textbf{h}}.
\end{equation}   
Here $\hat{x}$ and $\hat{p}$ are canonical Hermitian quadrature observables; for a specified mode $\textbf{s}$, $\hat{x}_{\textbf{s}} = \frac{a_{\textbf{s}}+a_{\textbf{s}}^{\dagger}}{\sqrt{2}}$ and $\hat{p}_{\textbf{s}} = \frac{a_{\textbf{s}}-a_{\textbf{s}}^{\dagger}}{\sqrt{2}i}$ satisfying the commutation relations $[\hat{x}_{\textbf{s}},\hat{x}_{\textbf{s}'}]=[\hat{p}_{\textbf{s}},\hat{p}_{\textbf{s}'}]=0,~$ and $[\hat{x}_{\textbf{s}},\hat{p}_{\textbf{s}'}] = i\delta_{\textbf{s},\textbf{s}'}$~\cite{simon1994quantum}. Also note that the sum $\hat{p}_{\textbf{k}}+\hat{p}_{\textbf{h}}$, and difference $\hat{x}_{\textbf{k}}-\hat{x}_{\textbf{h}}$ can be measured simultaneously since their commutator vanishes. Using the correspondence between continuous variable states and Wigner functions, the Wigner phase space distribution of the (bosonic) particle-hole pair can be written as~\cite{pirandola2006quantum},
\begin{equation}
\mathcal{W}(x_{\textbf{k}},p_{\textbf{k}},x_{\textbf{h}},p_{\textbf{h}})\propto\delta(x_{\textbf{k}}-x_{\textbf{h}})\delta(p_{\textbf{k}}+p_{\textbf{h}}).\label{wigner}
\end{equation}
The state can be understood as a two mode coherent squeeze state with $\zeta\rightarrow\infty$~\cite{pirandola2006quantum} (assuming $\zeta$ to be real, and $\phi=0$):
\begin{eqnarray}
	|\psi_{\textbf{k-h}}\rangle &=& \lim_{\zeta\rightarrow\infty}e^{\zeta a_{\textbf{k}}^{\dagger}a_{\textbf{h}}^{\dagger}-\zeta^{*}a_{\textbf{k}}a_{\textbf{h}}}|00\rangle\nonumber\\&=&\lim_{\zeta\rightarrow\infty}\text{sech}(\zeta)\sum_{m=0}^{\infty}\tanh(\zeta)^{m}|mm\rangle.
\end{eqnarray}
in agreement with our previous description. This limit is discussed more in the next section, in the context of an evaporating black hole.  

We emphasize that the interactions mediating this dynamical transfer of quantum information at normal fluid/superfluid BEC interface are strictly local. They are either the tunneling interaction at the interface which creates the quasi-particle excitations across the normal fluid/superfluid BEC boundary, or the scattering/ pairing interactions inside the condensate. This rules out the possibility of superluminal transfer of quantum information happening at the interface. In fact the speed of quantum information transfer can be roughly associated to the local speed of sound ($v_{c}$) in the superfluid state~\cite{manikandan2017andreev}. 
\section{The quantum final state of bosons falling into a black hole\label{sec3}}
In the previous section, we have discussed how the superfluid quantum ground state of interacting bosons can act like a fixed point in Hilbert space, such that a unitary description of quasiparticles interacting with the superfluid BEC is possible without changing the wavefunction of the superfluid BEC, when the condensate interactions are maximally entangling. Effectively, these processes can be described by considering the BEC necessitating a final state boundary condition for the modes which enter/leave the condensate. This, in turn preserves the quantum information encoded in the infalling modes via mode conversion processes which locally transfer the information to the outgoing modes.

We now look at the implications of this result for a black hole in its final stages of the evaporation process~\cite{horowitz2004black,lloyd2014unitarity,hawking1975particle,hayden2007black,susskind1993stretched,chakraborty2017black}. Our description of bosonic Andreev reflection from a normal fluid/superfluid BEC interface is quite similar to the ideas proposed by  Horowtiz and Maldacena~\cite{horowitz2004black}, and Preskill and Hayden~\cite{hayden2007black} respectively, regarding how the quantum information encoded in the infalling matter can escape from a black hole past the ``half-way'' point in the evaporation process, where more that half of its initial entropy has been radiated away and a quantum final state description can be envisaged. The prevalent idea these models propose is that a black hole in its final state, while accepting particles, acts like an ``information mirror"~\cite{hayden2007black} to the quantum information encoded in the infalling particles, and reflects the quantum information to the exterior via outgoing Hawking modes. In the following, we revisit the quantum information dynamics in these two models in the context of quantum information encoded in (continuous variable) bosonic modes and compare to the dynamics of a normal fluid/superfluid BEC interface.
\subsection{Analogy to the Horowitz-Maldacena model}
Following the condensate description provided in Sec.~\ref{sec1}, here we first describe how the information dynamics at the interface between a normal fluid and a superfluid BEC makes the condensate wavefunction analogous to Horowitz-Maldacena prescription for the quantum final state of a black hole~\cite{horowitz2004black}. In comparison to the original proposal by Horowitz and Maldacena, we consider quantum information encoded in continuous variable modes as appropriate for bosons~\cite{vaidman1994teleportation,pirandola2006quantum}. The resource for teleporting the quantum information to the exterior of a black hole is the entanglement between the particle/antiparticle pair produced at the event horizon (bosonic particle/hole pairs available at the interface between a normal fluid/superfluid BEC interface). We consider the limit $r,\zeta\rightarrow \infty$, corresponding to infinitely squeezed (maximally entangled) modes. A finite amount of squeezing in the final state would be analogous to the example described in Sec.~\ref{sec1}, where a departure from unitarity can be measured owing to not-maximally entangling interactions in the final state. This departure from unitarity in the final state projection approach has been discussed previously by  Gottesman and Preskill in Ref.~\cite{gottesman2004comment}. 

Since the Hawking pair is produced spontaneously from the vacuum, they are maximally entangled, and exist in time reversed states of each other such that they satisfy perfect correlations,
\begin{equation}
\hat{p}_{o}=-\hat{p}_{i}~~\text{and}~~\hat{x}_{o}=\hat{x}_{i},\label{maxE}
\end{equation}
in agreement with Eq.~\eqref{wigner}. The subscripts $i$ and $o$ denote ingoing and outgoing Hawking modes respectively.  
Now consider an incoming particle-like mode, described by the quadratures $\hat{x}_{in}$ and $\hat{p}_{in}$. Following Horowitz-Maldacena proposal for black hole evaporation, the infalling matter falls into the black hole (superfluid BEC) with the negative energy particle produced from vacuum fluctuations near the event horizon (fluctuations of the condensate at the interface between the superfluid/normal region), and the partner Hawking mode escapes to infinity. In the process, the black hole (in its final stages of the evaporation process) applies a final state boundary condition where the infalling modes are projected onto the relative position quadrature ($\hat{x}_{in}-\hat{x}_{i}$) basis, yielding result $x_{-}$, and to the sum of momentum quadratures ($\hat{p}_{in}+\hat{p}_{i}$) basis, yielding result $p_{+}$. Note that one can, in principle, simultaneously measure these two observables since they commute. As a result of perfect correlations described by Eq.~\eqref{maxE}, the outgoing antiparticle Hawking quanta has the spatial quadrature $\hat{x}_{o}=\hat{x}_{i}=\hat{x}_{in}-x_{-}$, and the momentum quadrature $\hat{p}_{o}=-\hat{p}_{i}=\hat{p}_{in}-p_{+}$~\cite{pirandola2006quantum,vaidman1994teleportation}. In the Schrodinger picture, if the wavefunction of the incoming particle was $\psi_{in}(x_{in})$, the wavefunction of the outgoing hole is unitarily related by spacial and momentum displacements, $\psi_{o}(x_{o})=e^{ix_{o}p_{+}}\psi_{in}(x_{o}+x_{-})$~\cite{vaidman1994teleportation}. 
\begin{figure}		\includegraphics[scale=1]{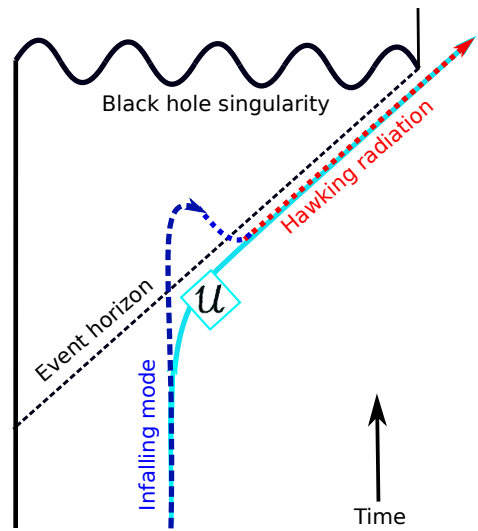}	\caption{Quantum information dynamics in Horowitz-Maldacena final state projection model, shown in the Penrose diagram of an evaporating black hole~\cite{horowitz2004black,lloyd2014unitarity,penrose2011republication,manikandan2017andreev}. It is shown that the quantum information encoded in the infalling mode is dynamically transferred to the outgoing Hawking radiation by the black hole imposing a particular final state boundary condition, where the unitarity of the process is described by the scattering matrix $\mathcal{U}$.\label{fig4}}
		\end{figure}

We now look at the case when the underlying space-time metric has spherical symmetry, and then it is natural to expect that the final quantum state of pairs of bosons entering into the black hole also respect this symmetry, which puts restrictions on the class of all possible candidate wavefunctions to describe the final quantum state of a black hole. The superfluid ground state of interacting bosons satisfy this symmetry: the two pairing modes are entangled in opposite momentum quantum states, where the paired wavefunction is symmetric under $\textbf{k}\rightarrow-\textbf{k}$, and the squeezing parameter depends only on the magnitude $|\textbf{k}|$, thereby strictly imposing the \textit{s-wave} symmetry for the infalling modes. 

We therefore conjecture that the final quantum state of bosons collapsing into a black hole is the maximally correlated superfluid quantum ground state of interacting bosons corresponding to the limit $r\rightarrow\infty$, $x_{-}=p_{+}=0$, analogous to Horowitz-Maldacena proposal~\cite{horowitz2004black}. The wavefunction of the outgoing mode in this case becomes $\psi_{o}(x_{o})=e^{ix_{o}p_{+}}\psi_{in}(x_{o}+x_{-})=\psi_{in}(x_{o})$~\cite{vaidman1994teleportation}, which is same as the wavefunction of the infalling mode. Here we considered the case when the phase of the superfluid is set to zero for simplicity. Assuming nonzero phase $\phi$ for the superfluid quantum ground state of maximally correlated pairs corresponds to a unitary transformation $\hat{U}=e^{-i\hat{N}\phi}$ relating the incoming and outgoing modes, as is evident from Eq.~\eqref{StTrans}. An illustration of the Horowitz-Maldacena proposal superimposed on the Penrose diagram of an evaporating black hole is shown in Fig.~\ref{fig4}.

Further, notice that the superfluid quantum ground state wavefunction, being macroscopic, is also rigid in some sense, and can act like a fixed point in the Hilbert space, where the dynamics can still respect unitarity and linearly of quantum physics without changing the wavefunction of the condensate. In laboratory settings, external parameters like temperature and pressure control give this stability to the condensate ground state, and we assume the boundary processes are also sufficiently low energy ($\varepsilon\rightarrow 0$) not to perturb the condensate from its ground state. This observation should be read in the context of an earlier argument made by Prof. Susskind, as to how, in the gravity context, requiring validity of a local quantum field theory respecting linearity and unitarity of quantum physics makes the wavefunction of the interior of a black hole independent of the past and future Cauchy surfaces~\cite{susskind1993stretched,chakraborty2017black}. By satisfying certain symmetry requirements as discussed above, we find that the superfluid ground state of interacting bosons can qualify as this special quantum state to describe the final quantum state of bosons collapsing into a black hole past its ``half-way" point in evaporation. 
\subsection{Black hole as an information mirror: analogy to Hayden-Preskill model\label{preskill}}
The ``information mirror" hypothesis for black hole evaporation was proposed by Hayden and Preskill to describe the final stages of an evaporating black hole, where the quantum information encoded in the infalling modes are rapidly revealed in the outgoing radiation~\cite{hayden2007black}. Here we resort to a minimal description of the Hayden-Preskill model proposed by Seth Lloyd and John Preskill~\cite{lloyd2014unitarity,manikandan2017andreev}. We show that a black hole assuming a superfluid quantum final state description in its late stages of the evaporation process can `Andreev' reflect the quantum information out of the superfluid while accepting particles, and therefore show that the normal fluid/superfluid BEC interface provide an experimentally realizable paradigm where the black hole-information mirror hypothesis can be investigated. 

We proceed as follows. Consider two quasiparticle excitation pairs at the boundary between a normal fluid and a superfluid BEC described by squeezing parameters $\zeta_{1}$ and $\zeta_{2}$ respectively. While there are quasiparticle excitation pairs created by the fluctuations of the BEC, we can also introduce these quasiparticle excitations externally. The two mode squeezed state $|\psi_{1}\rangle = \hat{S}(\zeta_{1})|0_{\textbf{m}}\rangle|0_{\textbf{q}}\rangle_{d}$ is assumed to describe a correlated information/external memory system, where the quantum information contained in the mode $\textbf{k}$ is described as correlations with an external memory system $\textbf{m}$. The second quasiparticle pair $|\psi_{2}\rangle = \hat{S}(\zeta_{2})|0_{\textbf{k}}\rangle_{d}|0_{\textbf{-k}}\rangle_{a}$ describes a bosonic particle/hole pair similar to our description in Sec.~\ref{sec1}.  Here we assume $\zeta_{1},~\zeta_{2}$ to be real for simplicity.
        
\begin{figure}		\includegraphics[scale=0.23]{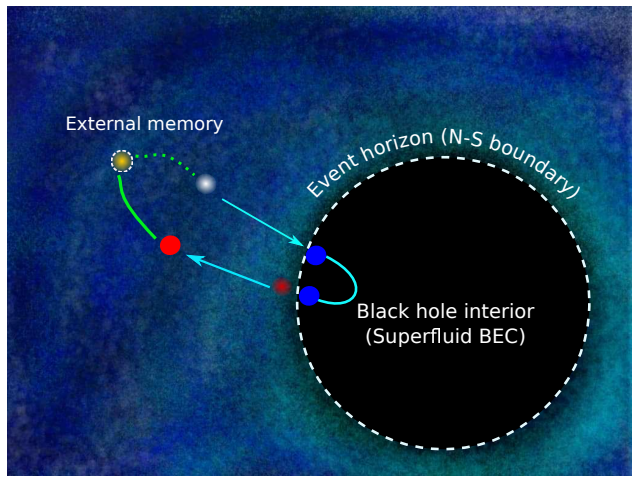}	\caption{Information mirror like processes near the event horizon[Normal fluid-Superfulid (N-S) boundary] following Hayden and Preskill's proposal ~\cite{hayden2007black,lloyd2014unitarity,manikandan2017andreev}: The incoming quasiparticle mode intially entangled with an external memory falls into the black hole by taking one particle from the Hawking pair, and forming a two mode paired entangled state inside the event horizon, while the outgoing Hawking mode, now correlated with the external memory, escapes to infinity.\label{fig3}}
		\end{figure}

When the mode $\textbf{k}$ is incident on the superfluid BEC interface from a normal fluid, the BEC require that the infalling particles form a particular entangled state inside the BEC, mediated by the condensate interactions. We implement this as the condensate applying a particular final state boundary condition for the infalling modes. The limiting case $\delta \textbf{k}\rightarrow 0$ is shown below, which corresponds to modes which can be present both in the condensate and the normal fluid. We find that applying the condensate boundary condition leaves the outgoing mode to be correlated with the memory system,
\begin{eqnarray}
&&|\psi_{\textbf{m},a}\rangle=\langle\psi(r,\textbf{k}_{\mu})|\psi_{1}\rangle\otimes|\psi_{2}\rangle\nonumber\\&&=\langle\psi(r,\textbf{k}_{\mu})|\hat{S}(\zeta_{1})|0_{\textbf{m}}\rangle|0_{\textbf{k}_{\mu}}\rangle_{d}\otimes\hat{S}(\zeta_{2})|0_{-\textbf{k}_{\mu}}\rangle_{d}|0_{\textbf{k}_{\mu}}\rangle_{a}\nonumber\\&&=\text{sech}(r)\sum_{n''=0}^{\infty}[e^{-i\phi}\tanh(r)]^{n''}\langle n''_{\textbf{k}_{\mu}}|_{d}\langle n''_{-\textbf{k}_{\mu}}|_{d}\nonumber\\&&~\text{sech}(\zeta_{1})\sum_{n'=0}^{\infty}\tanh(\zeta_{1})^{n'}|n'_{\textbf{m}}\rangle|n'_{\textbf{k}_{\mu}}\rangle_{d}\otimes\text{sech}(\zeta_{2})\nonumber\\&&\times\sum_{n=0}^{\infty}\tanh(\zeta_{2})^{n}|n_{-\textbf{k}_{\mu}}\rangle_{d}|n_{\textbf{k}_{\mu}}\rangle_{a}\propto\text{sech}(r)\text{sech}(\zeta_{1})\text{sech}(\zeta_{2})\nonumber\\&&\times\sum_{n=0}^{\infty}[e^{-i\phi}\tanh(r)\tanh(\zeta_{1})\tanh(\zeta_{2})]^{n}|n_{\textbf{m}}\rangle|n_{-\textbf{k}_{\mu}}\rangle_{a},
\end{eqnarray}
which describes entanglement between the memory system and the outgoing mode. Further note that in the limit $r,~\zeta_{1}~\text{and}~\zeta_{2}\rightarrow\infty$, we obtain perfect transfer of correlations. This can be seen immediately using the Heisenberg picture again. A perfectly correlated memory $\textbf{m}$ and the infalling mode $\textbf{q}$ is described by the correlations $\hat{x}_{\textbf{m}}=\hat{x}_{\textbf{q}}$ and $\hat{p}_{\textbf{m}}=-\hat{p}_{\textbf{q}}$ [see Eq.~\eqref{wigner}]. Another maximally correlated quasiparticle excitation pair (labeled by wavevector $\textbf{k}$) is produced due to the tunneling interactions at the interface, satisfying correlations $\hat{x}_{\textbf{k}}=\hat{x}_{-\textbf{k}}$ and $\hat{p}_{\textbf{k}}=-\hat{p}_{-\textbf{k}}$. Following our previous discussions, a spherically symmetric black hole (superfluid BEC) imposes the boundary condition that the infalling modes are maximally correlated; it imposes the correlations   $\hat{x}_{\textbf{q}}=\hat{x}_{\textbf{k}}$ and $\hat{p}_{\textbf{q}}=-\hat{p}_{\textbf{k}}$. As a consequence, we find that the outgoing matter is now maximally correlated to the memory system, $\hat{x}_{\textbf{m}}=\hat{x}_{-\textbf{k}}$ and $\hat{p}_{\textbf{m}}=-\hat{p}_{-\textbf{k}}$. In the sense of Preskill and Hayden, this describes a dynamical transfer of information from the infalling mode to the outgoing modes facilitated by the black hole applying an appropriate boundary condition for the infalling modes. Please see Fig.~\ref{fig3} for an illustration of this proposal.
\subsection{The effective temperature of Andreev processes}
During Andreev processes, an incident quasiparticle penetrates a finite distance into the condensate as evanescent waves~\cite{zapata2009andreev,klapwijk2004proximity}. For a superconductor, this length scale can be associated to the superconducting coherence length of the condensate~\cite{klapwijk2004proximity}. Similarly, for interacting bosons,we note that the corresponding length scale can be associated to the healing length of the condensate, $\lambda = \hbar/(\sqrt{2} m v_{c})$.  

This can be understood as follows: Andreev processes are most probable when the quasiparticles have roughly the energy equal to the chemical potential of the condensate, $\mu$. By equating $\frac{p^{2}}{2m}=\mu=mv_{c}^2$, we obtain the corresponding de Broglie wavelength, $\lambda = \hbar/(\sqrt{2} m v_{c})$. Note that this is essentially equivalent to equating the quantum pressure and interaction terms in the Hamiltonian as discussed in~\cite{dalfovo1999theory}. At the chemical potential, the Bogoliubov dispersion relation transition between phononic (linear dispersion) to free particle dispersion relation~\cite{andersen2004theory,kevrekidis2008basic}, therefore facilitating mode conversion processes mediated by condensate interactions.

From the most probable energy $\sim\mu$, Andreev processes can be associated to an effective temperature for the superfluid state,
\begin{equation}
T_{\lambda} = \frac{\hbar v_{c}}{\sqrt{2}k_{B}\lambda},
\end{equation}
where $v_{c}$ is the speed of sound in the superfluid state, and $\lambda$ is the healing length~\cite{macher2009black}. This is similar to the effective temperature of the superconducting ground state, which varies as $T_{\lambda_{s}}\sim \frac{\hbar v_{F}}{k_{B}\lambda_{s}}$ where $v_{F}$ is the Fermi velocity $\lambda_{s}$ is the coherence length of the superconductor~\cite{puspus2014entanglement,manikandan2017andreev}. By comparing the effective temperature of the superfluid state with the temperature of a Schwarzschild black hole~\cite{bekenstein1973black,jacobson1996introductory,ross2005black,hawking1975particle,manikandan2017andreev},
\begin{equation}
T_{BH}=\frac{\hbar c}{4\pi k_{B}r_{S}},
\end{equation}
where $c$ is the speed of light and $r_{S}$ the Schwarzschild radius, we arrive at the conclusion that in the Andreev analogy, the speed of sound is analogous to the speed of light, and the healing length of the superfluid is analogous to the Schwarzschild radius of a black hole.  
\section{Conclusions\label{conclusion}}
In this article, we proposed an analogy to investigate the resolutions of quantum information paradox in black holes by comparing the final quantum state of a black hole in its late stages of the evaporation process, to the superfluid quantum ground state of interacting bosons. We showed that in this scenario, the problem of particle production near the event horizon (pioneered by Prof. Stephen Hawking~\cite{hawking1975particle}) can be compared to mode conversion processes occurring at a normal fluid/ superfluid BEC interface, in such a way that the quantum information is preserved. Here, the quantum correlations encoded in continuous variable modes falling into a superfluid BEC can be dynamically transfered to the outgoing modes, facilitated by the superfluid applying a final state boundary condition to the infalling modes. 

In our analogy, the interface between a normal fluid and a superfluid BEC played the role of the event horizon in black holes, and the bulk of the superfluid was compared to the interior of a black hole where the quantum final state projection is applied. We further conjectured that the final quantum state for bosons falling into a black hole is the pair wavefunction of interacting bosons in a superfluid BEC ground state. The appeal of this conjecture is that (1) it resolves the information paradox in black holes by providing a unitary description of the particle production problem near the event horizon, (2) allows one to test quantum theories of gravity in controlled laboratory settings using normal fluid/superfluid BEC interfaces, and (3) in conjunction with the earlier result of the authors~\cite{manikandan2017andreev} where we proposed superconducting BCS wavefunction for fermions collapsing into a black hole, the bosonic analogy presented here, proposing the superfluid ground state of interacting bosons as the quantum final state of bosons falling into a black hole, gives a complete symmetric picture, prescribing quantum final state to both fermions and bosons inside a black hole. 

We stress that although we find surprisingly good similarity between the superfluid quantum ground state of interacting bosons (present study)/interacting fermions (see Ref.~\cite{manikandan2017andreev}), and the proposed quantum final state of a black hole in Horowitz-Maldacena, and Hayden-Preskill information mirror models, there are obvious differences between the two systems one can point out, and hence this analogy, considered alone, does not imply an exact correspondence between the two fields. What is remarkable is that albeit being a very complex many body system, a `classical' black hole is characterized by very few parameters, its mass, charge and angular momentum. Hence one would naively expect a microscopic quantum treatment for the final quantum state of black holes should also have similar simple set of parameters describing the dynamics, supporting our analogy to superfluid wavefunctions as candidate wavefunctions to describe the quantum final state of a black hole. 
\section{Acknowledgements}
We thank Prof. S. G. Rajeev for helpful discussions and suggestions. This work was supported by the John Templeton Foundation Grant ID 58558 and the National Science Foundation grants No. DMR-1809343.

\bibliography{bhref}
   \end{document}